\documentclass[pre,preprint,superscriptaddress,amsmath,amssymb,floatfix]
{revtex4}
\usepackage{graphics}
\usepackage[dvips]{graphicx}
\begin{document}
\title{A mesoscopic field theory of ionic systems versus a collective variable approach}
 \author{O. Patsahan}
\affiliation{Institute for Condensed Matter Physics of the National
Academy of Sciences of Ukraine, 1 Svientsitskii Str., 79011 Lviv, Ukraine}
 \author{ I. Mryglod}
\affiliation{Institute for Condensed Matter Physics of the National
Academy of Sciences of Ukraine, 1 Svientsitskii Str., 79011 Lviv, Ukraine}
 \date{\today}
\begin{abstract}
We establish a link between the two functional approaches: a mesoscopic field theory developed  recently by A.Ciach and G.Stell [A. Ciach and G. Stell, J. Mol. Liq. 87   (2000)  253] for  the study of ionic models and an exact statistical field theory based on the method of collective variables.
\end{abstract}
\maketitle

 Two rigorous scalar field theories were recently developed to describe the phase equilibria in ionic fluids:  the KSSHE (Kac-Siegert-Stratonovich-Hubbard-Ed\-wards) theory \cite{Cai-JSP} and the approach
 \cite {patsaha_mryglod,patsaha_mryglod_patsahan:06}  which is based on the collective variable (CV) method \cite{Zubar,Yuk1}. As was shown recently \cite{patsaha_mryglod_patsahan:06,caillol_patsahan_mryglod} both  of theories  are  in  close relation. Our goal here is to establish a link between the CV approach and the mesoscopic field theory developed in \cite{ciach:00:0,ciach:05:0} for the restricted primitive model with additional short-range interactions presented (RPM+SR) or, more specifically, to demonstrate how after some approximations in the exact microscopic CV action ${\mathcal H}[\nu_{\alpha},\rho,Q,\omega,\gamma]$ we can arrive at the functional of the grand potential
 $\Delta\Omega^{MF}[\eta,\phi]$ considered in \cite{ciach:00:0,ciach:05:0}.

Let us consider a general case of a classical two-component system consisting of $N$ particles among which there exist $N_{1}$ particles of species $1$ and $N_{2}$ particles of species $2$. The pair interaction potential is assumed to be of the following form:
\begin{equation}
U_{\alpha\beta}(r)=v_{\alpha\beta}^{HS}(r)+v_{\alpha\beta}^{C}(r)+v_{\alpha\beta}^{SR}(r),
\label{2.1a}
\end{equation}
where $v_{\alpha\beta}^{HS}(r)$ is the interaction potential between the two  additive hard spheres of diameters $\sigma_{\alpha\alpha}$ and $\sigma_{\beta\beta}$. We call the two-component hard sphere system a reference system (RS). Thermodynamic
and structural properties of RS are assumed to be known.
 $v_{\alpha\beta}^{C}(r)$ is the Coulomb potential: $v_{\alpha\beta}^{C}(r)=q_{\alpha}q_{\beta}v^{C}(r)$, where $v^{C}(r)=1/(D r)$, $D$ is the dielectric constant,  hereafter we put $D=1$. The solution is made of both positive and negative ions so that the electroneutrality is satisfied,
$\sum_{\alpha=1}^{2}q_{\alpha}c_{\alpha}=0$,
and $c_{\alpha}$ is the concentration of the species $\alpha$, $c_{\alpha}=N_{\alpha}/N$.
The ions of the species $\alpha=1$ are characterized by their hard sphere diameter $\sigma_{11}$ and their electrostatic charge $+q_{0}$ and those of species $\alpha=2$, characterized by diameter $\sigma_{22}$, bear opposite charge $-zq_{0}$ ($q_{0}$ is elementary charge and $z$ is the parameter of charge asymmetry).
  $v_{\alpha\beta}^{SR}(r)$ is the potential of the short-range interaction:
$v_{\alpha\beta}^{SR}(r)=v_{\alpha\beta}^{R}(r)+v_{\alpha\beta}^{A}(r)$, where $v_{\alpha\beta}^{R}(r)$ is used to mimic the soft core asymmetric repulsive interaction, $v_{\alpha\beta}^{R}(r)$ is assumed to have a Fourier transform; $v_{\alpha\beta}^{A}(r)$ describes a van der Waals-like attraction.

We consider the grand partition function (GPF) of the system which can be written as follows:
\begin{equation}
\Xi[\nu_{\alpha}]=\sum_{N_{1}\geq 0}\sum_{N_{2}\geq
0}\prod_{\alpha=1,2}
\frac{\exp(\nu_{\alpha}N_{\alpha})}{N_{\alpha}!} \int({\rm
d}\Gamma) \exp\left[-\frac{\beta}{2}\sum_{\alpha\beta}\sum_{ij}
U_{\alpha\beta}(r_{ij})\right].
\label{2.1}
\end{equation}
Here the following notations are used:
$\nu_{\alpha}$ is the dimensionless chemical potential, $\nu_{\alpha}=\beta\mu_{\alpha}-3\ln\Lambda$, $\mu_{\alpha}$ is the chemical potential of the $\alpha$th species, $\beta$ is the reciprocal temperature,
$\Lambda^{-1}=(2\pi m_{\alpha}\beta^{-1}/h^{2})^{1/2}$ is the inverse de Broglie thermal wavelength; $(\rm d\Gamma)$ is the element of configurational space of the particles.

Let us introduce operators $\hat\rho_{{\mathbf k}}$ and $\hat Q_{{\mathbf k}}$:
$\hat\rho_{{\mathbf k}}=\sum_{\alpha}\hat\rho_{{\mathbf k},\alpha}$ and $\hat Q_{{\mathbf
k}}=\sum_{\alpha}q_{\alpha}\hat\rho_{{\mathbf k},\alpha}$,
which are  combinations of the Fourier transforms of the microscopic number density of the species  $\alpha$:  $\hat\rho_{{\mathbf k},\alpha}=\sum_{i}\exp(-{\rm i}{\mathbf
k}{\mathbf r}_{i}^{\alpha})$. In this case  a part of the Boltzmann factor in (\ref{2.1}) which does not include the RS interaction can be presented as follows:
\begin{eqnarray}
&&\exp\left[-\frac{\beta}{2}\sum_{\alpha\beta}\sum_{i,j}(U_{\alpha\beta}(r_{ij})-v_{\alpha\beta}^{HS}
(r_{ij}))\right]=\exp\left[-\frac{1}{2}\sum_{{\bf k}}(\widetilde\Phi_{NN}\hat\rho_{{\mathbf k}}\hat\rho_{{\mathbf -k}}\right.  \nonumber \\
&&
\left.+\widetilde\Phi_{QQ}\hat Q_{{\mathbf k}}\hat Q_{{\mathbf -k}}+ 2\widetilde\Phi_{NQ}\hat\rho_{{\mathbf k}}\hat Q_{{\mathbf -k}})+\frac{1}{2}\sum_{\alpha}N_{\alpha}\sum_{{\mathbf
k}}(\widetilde\Phi_{\alpha\alpha}^{SR}(k)+q_{\alpha}^{2}\widetilde\Phi^{C}(k))\right],
\label{2.2}
\end{eqnarray}
where
\begin{eqnarray}
\widetilde{\Phi}_{NN}(k)&=&\frac{1}{(1+z)^{2}}\left[
z^{2}\widetilde{\Phi}_{11}^{SR}(k) + 2z \widetilde{\Phi}_{12}^{SR}(k)
+\widetilde{\Phi}_{22}^{SR}(k)\right],  \nonumber \\
\widetilde{\Phi}_{QQ}(k)&=&\frac{1}{(1+z)^{2}}\left[
\widetilde{\Phi}_{11}^{SR}(k)
-2\widetilde{\Phi}_{12}^{SR}(k)+\widetilde{\Phi}_{22}^{SR}(k)\right]
+\widetilde{\Phi}^{C}(k), \nonumber \\
\widetilde{\Phi}_{NQ}(k)&=&\frac{1}{(1+z)^{2}}\left[ z
\widetilde{\Phi}_{11}^{SR}(k)
+(1-z)\widetilde{\Phi}_{12}^{SR}(k)-\widetilde{\Phi}_{22}^{SR}(k)\right]
\label{2.3}
\end{eqnarray}
and we use the notations $\widetilde{\Phi}_{\alpha\beta}^{X\ldots}(k)=\frac{\beta}{V}\widetilde v_{\alpha\beta}^{X\ldots}(k)$ with
$\widetilde{v}_{\alpha\beta}^{X\ldots}(k)$ being a Fourier transform of the corresponding interaction potential.

In order to introduce the collective variables (CVs) we use  the identity
\begin{equation}
\exp\left[-\frac{1}{2}\sum_{{\bf k}}\widetilde\Phi(k)\hat\xi_{{\mathbf k}}\hat\xi_{-{\mathbf k}}\right] =\int({\rm
d}\xi)\delta_{{\mathcal
F}}[\xi-\hat\xi]\exp\left[-\frac{1}{2}\sum_{{\bf k}}\widetilde\Phi(k)\xi_{{\mathbf k}}\xi_{-{\mathbf k}}\right],
\label{2.4}
\end{equation}
where $\delta_{{\mathcal F}}[\xi-\hat\xi]$ denotes the functional delta function.

Taking into account (\ref{2.2})-(\ref{2.4}),  we can rewrite (\ref{2.1})
\begin{equation}
\Xi[\nu_{\alpha}]=\int ({\rm d}\rho)({\rm d}Q)({\rm
d}\omega)({\rm d}\gamma) \exp\left(-{\mathcal H}[\nu_{\alpha},\rho,Q,\omega,\gamma] \right),
\label{2.5}
\end{equation}
where the CV action ${\mathcal H}$ is as follows:
\begin{eqnarray}
{\mathcal H}[\nu_{\alpha},\rho,Q,\omega,\gamma]&=&\frac{1}{2}\sum_{{\mathbf
k}}[\widetilde \Phi_{NN}(k)\rho_{{\mathbf k}}\rho_{-{\mathbf
k}}+\widetilde \Phi_{QQ}(k)Q_{{\mathbf k}}Q_{-{\mathbf k}}+2\widetilde \Phi_{NQ}(k)\rho_{{\mathbf k}}Q_{-{\mathbf
k}}] \nonumber\\
&&
-{\rm i}\sum_{{\mathbf k}}(\omega_{{\mathbf k}}\rho_{{\mathbf
k}}+\gamma_{{\mathbf k}}Q_{{\mathbf k}})-\ln \Xi_{HS}[\bar
\nu_{\alpha};-{\rm i}\omega,-{\rm i}q_{\alpha}\gamma].
\label{2.5a}
\end{eqnarray}
In (\ref{2.5a}) CVs $\rho_{{\mathbf k}}$ and $Q_{{\mathbf k}}$ describe fluctuations of the total number density and charge density, respectively.
$\Xi_{HS}[\bar\nu_{\alpha};-{\rm i}\omega,-{\rm i}q_{\alpha}\gamma]$ is the GPF of a two-component system of  hard spheres with the renormalized local chemical potential
\begin{equation}
\bar \nu_{\alpha}=\nu_{\alpha}+\frac{1}{2}\sum_{{\mathbf
k}}\widetilde\Phi_{\alpha\alpha}^{SR}(k)+\frac{q_{\alpha}^{2}}{2}\sum_{{\mathbf
k}}\widetilde\Phi^{C}(k)-{\rm i}\omega(r)+{\rm
i}q_{\alpha}\gamma(r).
\label{2.7}
\end{equation}

 The MF approximation of  functional (\ref{2.5}) is defined by
\begin{equation}
\Xi_{MF}[\nu_{\alpha}]=\exp(-{\mathcal{H}}[\nu_{\alpha},\bar\rho, \bar Q, \bar\omega, \bar\gamma]),
\label{2.9}
\end{equation}
where $\bar\rho$, $\bar Q$, $\bar\omega$ and $\bar\gamma$
are the solutions of the saddle point equations:
\begin{eqnarray}
\bar\rho &=&\langle N[\bar \nu_{\alpha};-{\rm i}\bar\omega,-{\rm
i}q_{\alpha}\bar\gamma]\rangle_{HS}, \qquad \bar Q=0, \nonumber \\
\bar\omega& =& -{\rm{i}}\bar\rho \widetilde{v}_{NN}(0), \qquad
\bar\gamma=-{\rm{i}} \bar\rho \widetilde{v}_{NQ}(0).
\label{2.10}
\end{eqnarray}

Now we present CVs $\rho_{{\mathbf k}}$ and  $Q_{{\mathbf k}}$
($\omega_{{\mathbf k}}$ and $\gamma_{{\mathbf
k}}$) as
\begin{eqnarray*}
\rho_{{\mathbf k}}&=&\bar\rho \delta_{{\mathbf
k}}+\delta\rho_{{\mathbf k}}, \qquad Q_{{\mathbf k}}=\bar Q
\delta_{{\mathbf k}}+\delta Q_{{\mathbf k}}, \nonumber \\
\omega_{{\mathbf k}}&=&\bar\omega \delta_{{\mathbf
k}}+\delta\omega_{{\mathbf k}}, \qquad \gamma_{{\mathbf
k}}=\bar\gamma \delta_{{\mathbf k}}+\delta\gamma_{{\mathbf k}},
\end{eqnarray*}
where the quantaties with a bar are given by (\ref{2.10}) and $\delta_{{\mathbf k}}$ is the Kronecker symbol.
Then we write  $\ln\Xi_{HS}[\bar\nu_{\alpha};-{\rm i}\omega,-{\rm i}q_{\alpha}\gamma]$   in the form of the cumulant expansion
\begin{eqnarray}
\ln\Xi_{HS}[\ldots]&=&\sum_{n\geq 0}\frac{(-{\rm
i})^{n}}{n!}\sum_{i_{n}\geq 0}
\sum_{{\mathbf{k}}_{1},\ldots,{\mathbf{k}}_{n}}
{\mathfrak{M}}_{n}^{(i_{n})}(k_{1},\ldots,k_{n})\delta\gamma_{{\mathbf{k}}_{1}}\ldots\delta
\gamma_{{\mathbf{k}}_{i_{n}}} \nonumber \\
&&\times
\delta\omega_{{\mathbf{k}}_{i_{n+1}}}\ldots\delta\omega_{{\mathbf{k}}_{n}}\delta_{{\mathbf{k}}_{1}+\ldots
+{\mathbf{k}}_{n}},
\label{2.11}
\end{eqnarray}
where ${\mathfrak{M}}_{n}^{(i_{n})}(k_{1},\ldots,k_{n})$ is the $n$th cumulant  defined by
\begin{equation}
{\mathfrak{M}}_{n}^{(i_{n})}(k_{1},\ldots,k_{n})=\frac{\partial^{n}\ln
\Xi_{HS}[\ldots]}{\partial
\delta\gamma_{{\bf{k}}_{1}}\ldots\partial\delta\gamma_{{\bf{k}}_{i_{n}}}
\partial\delta\omega_{{\bf{k}}_{i_{n+1}}}\ldots\partial\delta\omega_{{\bf{k}}_{n}}}\vert_{\delta\gamma_{{\mathbf{k}}} =0,\delta\omega_{{\mathbf{k}}}=0}.
\label{2.12}
\end{equation}
The expressions for the cumulants (for $n\leq 4$) are given in \cite{patsaha_mryglod_patsahan:06}.
Substituting (\ref{2.11})-(\ref{2.12}) in (\ref{2.5})-(\ref{2.5a}) we obtain
\begin{eqnarray}
\Xi[\nu_{\alpha}]&=&\Xi_{MF}[\bar\nu_{\alpha}] \int(\mathrm{d}\delta\rho)(\mathrm{d}\delta
Q)(\mathrm{d}\delta\omega)(\mathrm{d}\delta\gamma)\exp\Big\{-\frac{1}{2!}\sum_{\bf
k}\left[\widetilde \Phi_{NN}(k)\delta\rho_{\bf k}\delta\rho_{-\bf
k}\right. \nonumber \\
 &&\left.+2\widetilde \Phi_{NQ}(k)\delta\rho_{\bf k}\delta Q_{-\bf k}+\widetilde \Phi_{QQ}(k)\delta Q_{\bf k}\delta Q_{-\bf k}\right]
+{\rm i}\sum_{{\mathbf k}}(\delta\omega_{{\mathbf k}}\delta\rho_{{\mathbf
k}}+\delta\gamma_{{\mathbf k}}\delta Q_{{\mathbf k}}) \nonumber \\
 &&
+\sum_{n\geq 2}\frac{(-{\rm
i})^{n}}{n!}\sum_{i_{n}\geq 0}
\sum_{{\mathbf{k}}_{1},\ldots,{\mathbf{k}}_{n}}
{\mathfrak{M}}_{n}^{(i_{n})}(k_{1},\ldots,k_{n})\delta\gamma_{{\bf{k}}_{1}}\ldots\delta\gamma_{{\bf{k}}_{i_{n}}}
\nonumber \\
&&
\times\delta\omega_{{\bf{k}}_{i_{n+1}}}\ldots\delta\omega_{{\bf{k}}_{n}}\delta_{{\bf{k}}_{1}+\ldots
+{\bf{k}}_{n}}\Big\}.
\label{dA.14}
\end{eqnarray}
Let us make the following approximations. We neglect the $k$-dependence of the cumulants putting
${\mathfrak{M}}_{n}^{(i_{n})}(k_{1},\ldots,k_{n})\simeq {\mathfrak{M}}_{n}^{(i_{n})}(0,\ldots,0)$ and replace the full chemical potentials $\nu_{\alpha}$ by their MF values $\nu_{\alpha}^{*}$.
We also limit our consideration to the restricted primitive model ($z=1$) supplemented
by the same short-range interactions for both species (RPM+SR) ($v_{11}^{SR}(r)=v_{22}^{SR}(r)=v_{12}^{SR}(r)=v^{SR}(r)$). The latter means that $\widetilde{\Phi}_{NN}(k)=\widetilde{\Phi}^{SR}(k)$,  $\widetilde{\Phi}_{NQ}(k)=0$ and $\widetilde{\Phi}_{QQ}(k)=\widetilde{\Phi}^{C}(k)$. In this case we  have for the cumulants \cite{patsaha_mryglod}
\begin{eqnarray}
{\mathfrak{M}}_{n}^{(0)}&=&{\widetilde G}_{n}, \qquad
{\mathfrak{M}}_{n}^{(2)}=q_{0}^{2}{\widetilde G}_{n-1}, \qquad
{\mathfrak{M}}_{n}^{(3)}\equiv 0, \nonumber \\
{\mathfrak{M}}_{n}^{(4)}&=&q_{0}^{4}(3{\widetilde G}_{n-2}-2{\widetilde G}_{n-3}),
\label{cumulant}
\end{eqnarray}
  where ${\widetilde G}_{n}={\widetilde G}_{n}(0,\ldots,0)$ is the Fourier transform of the $n$-particle truncated (connected) correlation function \cite{stell} of a one-component hard sphere system with the density $\bar\rho$ defined by  (\ref{2.10}). Because both ${\mathfrak{M}}_{2}^{(0)}$ and ${\mathfrak{M}}_{2}^{(2)}$ are positive and smooth functions in the region under consideration
\cite{patsaha_mryglod}, we can integrate in (\ref{dA.14}) over $\delta\omega_{\bf{k}}$ and  $\delta\gamma_{\bf{k}}$ with the Gaussian density measure as basic one. The integration is performed using the Euler equations. We determine $\delta\omega_{\bf{k}}^{*}$ and $\delta\gamma_{\bf{k}}^{*}$ which provide a maximum for the functional in the exponent of (\ref{dA.14}). This leads to the  expression for $\Xi$
\begin{equation}
\Xi=\Xi_{MF}{\mathcal C}\int ({\rm d}\delta\rho)({\rm d}\delta Q)\exp\left(-\widetilde{\mathcal H}(\delta\rho,\delta Q) \right),
\label{3.5}
\end{equation}
where
\begin{eqnarray}
\widetilde{\mathcal H}(\delta\rho,\delta Q)&=&\frac{1}{2}\sum_{{\mathbf
k}}[a_{2}^{(0)}(k)\delta\rho_{{\mathbf k}}\delta\rho_{-{\mathbf
k}}+a_{2}^{(2)}(k)\delta Q_{{\mathbf k}}\delta Q_{-{\mathbf k}}]+\sum_{n\geq 3}\frac{1}{n!}\sum_{i_{n}\geq 0} \nonumber\\
&&
\times
\sum_{{\mathbf{k}}_{1},\ldots,{\mathbf{k}}_{n}}
a_{n}^{(i_{n})}\delta Q_{{\bf{k}}_{1}}\ldots\delta Q_{{\bf{k}}_{i_{n}}}\delta\rho_{{\bf{k}}_{i_{n+1}}}\ldots\delta\rho_{{\bf{k}}_{n}}\delta_{{\bf{k}}_{1}+\ldots
+{\bf{k}}_{n}}.
\label{3.5b}
\end{eqnarray}
and the following notations are introduced
\[
{\mathcal C}= \prod_{{\mathbf k}}\frac{1}{\pi{\mathfrak{M}}_{2}^{(0)}}\prod_{{\mathbf k}}\frac{1}{\pi{\mathfrak{M}}_{2}^{(2)}},
\]
\begin{equation}
a_{2}^{(0)}(k)=\widetilde{\Phi}^{SR}(k)+1/{\mathfrak{M}}_{2}^{(0)}, \qquad
 a_{2}^{(2)}(k)=\widetilde{\Phi}^{C}(k)+1/{\mathfrak{M}}_{2}^{(2)},
 \label{coef_1}
\end{equation}
\begin{equation}
 a_{3}^{(0)}=-{\mathfrak{M}}_{3}^{(0)}/\left({\mathfrak{M}}_{2}^{(0)}\right)^{3}, \qquad
a_{3}^{(2)}=-\frac{3{\mathfrak{M}}_{3}^{(2)}}{{\mathfrak{M}}_{2}^{(0)}\left( {\mathfrak{M}}_{2}^{(0)}\right) ^{2}},
\label{coef_1a}
\end{equation}
\begin{equation}
a_{4}^{(0)}=-\frac{1}{\left({\mathfrak{M}}_{2}^{(0)}\right)^{4}}\left[{\mathfrak{M}}_{4}^{(0)}-\frac{3\left( {\mathfrak{M}}_{3}^{(0)}\right)^{2}}{{\mathfrak{M}}_{2}^{(0)}}\right], \qquad
a_{4}^{(2)}=\frac{12\left( {\mathfrak{M}}_{3}^{(2)}\right)^{2}}{\left({\mathfrak{M}}_{2}^{(0)}\right)^{2}\left({\mathfrak{M}}_{2}^{(2)}\right)^{3}},
 \label{coef_2}
\end{equation}
\begin{equation}
a_{4}^{(4)}=-\frac{1}{\left({\mathfrak{M}}_{2}^{(2)}\right)^{4}}\left[{\mathfrak{M}}_{4}^{(4)}-\frac{3\left( {\mathfrak{M}}_{3}^{(2)}\right)^{2}}{{\mathfrak{M}}_{2}^{(0)}}\right].
 \label{coef_3}
\end{equation}
Taking into account (\ref{cumulant}) we rewrite (\ref{3.5b})-(\ref{coef_3}) as follows:
\begin{equation}
\Xi=\Xi_{MF}{\mathcal C}'\int ({\rm d}\delta\rho)({\rm d}\delta Q^{*})\exp\left(-\widetilde{\mathcal H}(\delta\rho,\delta Q^{*}) \right),
\label{3.6b}
\end{equation}
where
\begin{eqnarray}
\widetilde{\mathcal H}(\delta\rho,\delta Q^{*})&=&\frac{1}{2V}\sum_{{\mathbf
k}}[{\widetilde a}_{2}^{(0)}(k)\delta\rho_{{\mathbf k}}\delta\rho_{-{\mathbf
k}}+{\widetilde a}_{2}^{(2)}(k)\delta Q^{*}_{{\mathbf k}}\delta Q^{*}_{-{\mathbf k}}]+\sum_{n\geq 3}\frac{1}{n!}\frac{1}{V^{n-1}}\nonumber\\
&&
\times\sum_{i_{n}\geq 0}
\sum_{{\mathbf{k}}_{1},\ldots,{\mathbf{k}}_{n}}
{\widetilde a}_{n}^{(i_{n})} \delta Q^{*}_{{\bf{k}}_{1}}\ldots\delta Q^{*}_{{\bf{k}}_{i_{n}}}\delta\rho_{{\bf{k}}_{i_{n+1}}}\ldots\delta\rho_{{\bf{k}}_{n}}\delta_{{\bf{k}}_{1}+\ldots+{\bf{k}}_{n}}.
\label{3.5a}
\end{eqnarray}
Here
\[
{\widetilde a}_{2}^{(0)}(k)=\beta\widetilde{v}^{SR}(k)+\frac{1}{\bar\rho {\widetilde S}_{2}}, \qquad
{\widetilde a}_{2}^{(2)}(k)=\beta\widetilde{v}^{C}(k)+\frac{1}{\bar\rho},
\]
\[
{\widetilde a_{3}}^{(0)}=-\frac{{\widetilde S}_{3}}{\bar\rho^{2}{\widetilde S}_{2}^{3}}, \qquad
{\widetilde a}_{3}^{(2)}=-\frac{3}{\bar\rho^{2}},
 \label{tildecoef_1a}
\]
\[
{\widetilde a}_{4}^{(0)}=\frac{1}{\bar\rho{\widetilde S}_{2}^{4}}\left({\widetilde S}_{4}-3\frac{{\widetilde S}_{3}^{2}}{{\widetilde S}_{2}}\right),\qquad
{\widetilde a}_{4}^{(2)}=\frac{12}{\bar\rho^{3}}, \qquad
{\widetilde a}_{4}^{(4)}=\frac{2}{\bar\rho^{3}}
\]
with $\delta Q^{*}_{{\mathbf k}}=\delta Q_{{\mathbf k}}/q_{0}$ and ${\widetilde S}_{n}={\widetilde G}_{n}/\langle N\rangle$.

After taking into account $\frac{1}{V}\sum_{{\mathbf k}}=(2\pi)^{-3}\int{\rm d}{\mathbf k}$ and
 $\delta_{{\bf{k}}_{1}+\ldots+{\bf{k}}_{n}}=\frac{1}{V}\int{\rm d}{\mathbf r}\exp{\rm i}{\mathbf r}({\mathbf k_{1}}+\ldots+{\mathbf k}_{n})$ in (\ref{3.5a}) we arrive at the same expression for the action as that obtained in the mesoscopic field theory  (see e.g., Eqs. (44)-(48) in \cite{ciach:05:0}), $\Delta\Omega^{MF}[\tilde\eta,\tilde\phi]$  with  $\widetilde\eta({\bf{k}})=\delta\rho_{{\bf{k}}}$ and $\widetilde\phi({\bf{k}})= \delta Q^{*}_{{\bf{k}}}$. In order to demonstrate that the corresponding coefficients  of both actions coincide i.e.,
\begin{equation}
{\widetilde a}_{2}^{(0)}(k)={\widetilde C}_{\eta\eta}^{(0)}(k)=\beta\widetilde{v}^{SR}(k)+\gamma_{0,2},
\label{corresp1}
\end{equation}
\begin{equation}
{\widetilde a}_{2}^{(2)}(k)={\widetilde C}_{\phi\phi}^{(0)}(k)=\beta\widetilde{v}^{C}(k)+\gamma_{2,0},
\label{corresp2}
\end{equation}
\begin{equation}
{\widetilde a}_{3}^{(0)}=\gamma_{0,3}, \quad
{\widetilde a}_{3}^{(2)}=3\gamma_{2,1}, \quad
{\widetilde a}_{4}^{(0)}=\gamma_{0,4}, \quad
{\widetilde a}_{4}^{(2)}=6\gamma_{2,2}, \quad
{\widetilde a}_{4}^{(4)}=\gamma_{4,0}
\label{corresp3}
\end{equation}
(${\widetilde C}_{\eta\eta}^{(0)}(k)$, ${\widetilde C}_{\phi\phi}^{(0)}(k)$ and $\gamma_{2m,n}$ are the notations used in \cite{ciach:05:0}), we consider coefficients $\gamma_{2m,n}$ in detail. To this end we recall that   $\gamma_{2m,n}$ denotes the appropriate derivative of the Helmholtz free energy of the hard sphere system $f_{h}$ (see e.g.\cite{ciach:05:0})
\[
\gamma_{2m,n}=\beta\frac{\partial^{2m+n}f_{h}}{\partial \phi^{2m}\partial \rho^{*^{n}}}\vert_{\phi=0,\rho^{*}=\bar\rho^{*}}.
\]
For example, when the Carnahan-Starling (CS) form of $f_{h}$ is adobted in the local-density approximation
\begin{eqnarray*}
\beta f_h(\rho^*,\phi)=\frac{\rho^*+\phi}{2}
\log\Big(\frac{\rho^*+\phi}{2}\Big)
+\frac{\rho^*-\phi}{2}\log\Bigg(\frac{\rho^*-\phi}{2}\Bigg)-\rho^*
+\rho^{*}\frac{s(4-3s)}{(1-s)^{2}},
\end{eqnarray*}
we obtain the following explicit expressions for $\gamma_{2m,n}$
\begin{eqnarray}
\gamma_{0,2}&=&\frac{1+4s+4s^{2}-4s^{3}+s^{4}}{(1-s)^{4}\bar\rho^{*}}=\frac{1}{\bar\rho^{*}{\widetilde S}_{2}}, \quad
\gamma_{2,0}=\frac{1}{\bar\rho^{*}}
\label{gamma}
 \\
\gamma_{0,3}&=&-\frac{1-5s-20s^{2}-4s^{3} +5s^{4}-s^{5}}{\bar\rho^{*^{2}}(1-s)^{5}}=-\frac{{\widetilde S}_{3}}{\bar\rho^{*^{2}}{\widetilde S}_{2}^{3}}, \quad
\gamma_{2,1}=-\frac{1}{\bar\rho^{*^{2}}},
\label{gamma1}
 \\
\gamma_{0,4}&=&\frac{2(1-6s+15s^{2}+52s^{3}+3s^{4}-6s^{5}+s^{6})}{\bar\rho^{*^{3}}(1-s)^{6}}=
\frac{1}{\bar\rho^{*^{3}}{\widetilde S}_{2}^{4}}\left({\widetilde S}_{4}-3\frac{{\widetilde S}_{3}^{2}}{{\widetilde S}_{2}}\right),
\label{gamma2}
 \\
\gamma_{2,2}&=&\frac{2}{\bar\rho^{*^{3}}}, \qquad
\gamma_{4,4}=\frac{2}{\bar\rho^{*^{3}}}.
\label{gamma3}
\end{eqnarray}
In the above equations we use the notations $\rho^*=\bar\rho^{*}+\delta\rho^{*}$ and $s=\pi \rho^*/6$ ($\rho^{*}=\rho\sigma^{3}$ and $\bar\rho^{*}$ are the dimensionless total number density and its MF value respectively).
As is seen, expressions (\ref{gamma})-(\ref{gamma3})  confirm the relations given by (\ref{corresp1})-(\ref{corresp3}).

\end{document}